\documentclass[a4paper,fleqn,usenatbib]{mnras}

\usepackage[T1]{fontenc}
\usepackage{ae,aecompl}
\usepackage{lineno}
\usepackage{threeparttable}

\usepackage{graphicx}	
\usepackage{amssymb}	
 
\title[Degree of electric current neutralization and the activity in ARs]{Degree of electric current neutralization and the activity in solar Active Regions}

\author[P.~Vemareddy]{
P.~Vemareddy,$^{1}$\thanks{E-mail: vemareddy@iiap.res.in}
\\
$^{1}$Indian Institute of Astrophysics, Sarjapur road, II Block, Koramangala, Bengaluru-560 034, India
}
\date{Accepted April 5, 2019;  Received April 4, 2019; in original form February 10, 2019}

\pubyear{2019}

\begin{document}
\label{firstpage}
\pagerange{\pageref{firstpage}--\pageref{lastpage}}
\maketitle

\begin{abstract}
Using time-sequence vector magnetic field observation from \textit{Helioseismic and Magnetic Imager}, we examined the connection of non-neutralized currents and the observed activity in 20 solar active-regions (ARs). The net current in a given magnetic polarity is algebraic sum of direct-current (DC) and return-current (RC) and the ratio $|DC/RC|$ is a measure of degree of net-current-neutralization (NCN). In the emerging ARs, the non-neutrality of these currents builds with the onset of flux emergence, following the relaxation to neutrality during the separation motion of bipolar regions. Accordingly, some emerging ARs are source regions of CMEs occurring at the time of higher level non-neutrality. ARs in the post-emergence phase can be CME productive provided they have interacting bipolar regions with converging and shearing motions. In these cases, the net current evolves with higher level ($>1.3$) of non-neutrality. Differently, the $|DC/RC|$ in flaring and quiet ARs vary near unity. In all the AR samples, the $|DC/RC|$ is higher for chiral current density than that for vertical current density. Owing to the fact that the non-neutralized currents arise in the vicinity of sheared polarity-inversion-lines (SPILs), the profiles of the total length of SPIL segments and the degree of NCN follow each other with a positive correlation. We find that the SPIL is localized as small segments in flaring-ARs whereas it is long continuous in CME-producing ARs. These observations demonstrate the dividing line between the CMEs and flares with the difference being in global or local nature of magnetic shear in the AR that reflected in non-neutralized currents.
\end{abstract}




\section{Introduction}
\label{Intro}
It is widely accepted that solar eruptions including flares and CMEs are magnetically driven events and had been the subject of study for the past four decades. Mainly, these transient, eruptive events derive their energy from electrical currents that pass through the solar atmosphere. There are evidential reports that the strong electrical currents involved in major flares are embedded within magnetic flux emerged from convection zone through photosphere into the corona (e.g., \citealt{leka1996, vemareddy2012b}).  Other sources that generate these currents are plasma displacements on the surface of the sun \citep{leka2003a, leka2003b, schrijver2005, vemareddy2012b,vemareddy2015c}. Dissipation of these electrical currents during reconnection is the supposed mechanism of releasing magnetic energy to thermal and kinetic energy during the transient activity of the sun \citep{priest2002}.

However, as theorized by \citet{parker1996}, the net electric current at any cross-section of a flux tube embedded in relatively field-free plasma must result in a zero value. In this scenario, the net current of a flux tube is a sum of direct current due to its twist/shear and the return current in a interface layer between the flux tube and the surrounding plasma. These two (body and surface) currents are opposite in sign but equal in magnitude. As the flux tubes emanate from the photosphere, which are confined by the dense plasma, the integrated net vertical current from the flux tube cross section is expected to obey the above Parker's paradigm of neutralized current. However, the observations of net current over a polarity are quiet deviating from neutrality \citep{venkat2009, ravindra2011, georgoulis2012, vemareddy2015c} indicating controversial nature of theory and observations. 

In the presence of plasma flows, the field lines connecting the well-isolated polarities are stressed, nevertheless, the generated currents at their foot points remain neutralized, according to integral form of Ampere's law. Even emergence of current carrying flux tubes in the case of well-separated, carry vanishing currents. Indeed, using high-resolution observations of photospheric magnetic fields, it is found that the net-current from many isolated sunspots is near to neutrality \citep{venkat2009}. The question to ask what are the factors that contribute to the net current from a single polarity. \citet{georgoulis2012} had investigated these issues extensively in two ARs, where they concluded that intense polarity inversion lines (PILs) amid the compact polarity regions in the core of the ARs support significantly for the non-neutralized currents. This result is further supplemented by a recent study of \citet{vemareddy2015c}, where the length of strong magnetically sheared PIL (SPIL) was found to be proportional to the observed net current. In this context, a very early study by \citet{falconer2001} was the basis for this well proved relation according to which the presence of strong SPIL is an indicator for the occurrence of eruptions and its length is a measure of net vertical current. 

The plasma $\beta$ in the coronal environment is far less than unity yielding to gradually dominating field aligned currents over the return currents along the flux tube. As a result, there is a net current over the cross-section of the flux tube in the corona. Due to this stratification in plasma-$\beta$, it is not well understood, how the net current varies while emerging, especially at and after the instance of emergence. Employing numerical simulations of emerging bipolar AR, this issue was investigated by \citet{torok2014}. Neutralization was observed to increasingly break down during emergence and retained at a slightly lower value after emergence stopped. However, it has not yet well investigated whether this is a characteristic evolution in all emerging ARs. On the other hand, what is the scenario in the post-emerged ARs where a different boundary motions drive the evolution? In order to have more insights, the net current evolution and its characteristic time profile of neutralization needs to be systematically examined from observations of different ARs, which is the motivation of this study. 

Recently, systematic observational study of \citet{YangLiu2017} confirmed that the SPIL is the major contributor to DC and proposed the degree of current-neutralization as the proxy for assessing the ability of ARs to produce CMEs. In this view,  it would be important to study the the connection between the degree of net current-neutralization (NCN) with the nature of generated activity, viz., CMEs or flares, in a large sample of ARs.   

\begin{figure*}[!htp]
	\centering
	\includegraphics[width=.99\textwidth,clip=]{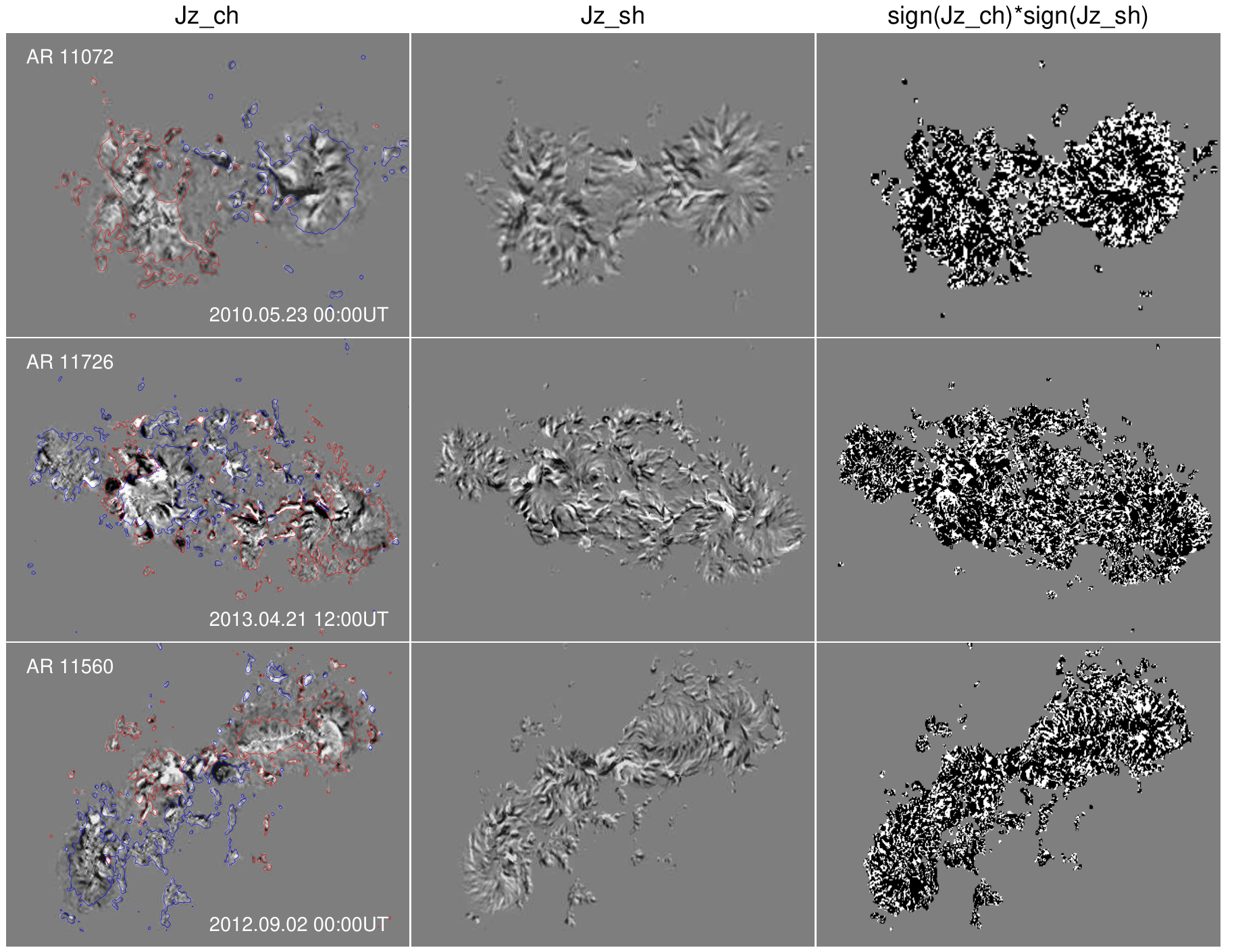}
	\caption{First column: Distribution of $Jz_{ch}$, with blue/red contours of $B_z$ at $\pm$150G,  second column: Distribution of $Jz_{sh}$, All maps are scaled within $\pm 50 mAm^{-2}$. third column: relative sign of $Jz_{ch}$ and $Jz_{sh}$. Black (white) pixels refer to those having opposite (same) sign of $Jz_{ch}$ and $Jz_{sh}$, which are 60-65\% (35-40\%) in all over the AR. }
	\label{Fig1}
\end{figure*}

Further, recent study by \citet{vemareddy2017d} shows that the gradient of magnetic field strength contributes to opposite signed, although smaller in magnitude, current to that contributed by magnetic field direction in the vertical component of current. As a result, the net current neutralization is enhanced in a polarity and could substantiates the origins of sheath current in field strength gradients as first indicated in \citet{zhangh2010}. This observational evidence also requires solid basis in more AR cases. The now available uninterrupted vector magnetic field observations from \textit{Helioseismic and Magnetic Imager} are very identical for the objectives of this study, to supplement possible scenarios of electric current evolution. The rest of the paper is organized as follows. The observational data and analysis procedure is detailed in Section~\ref{ObsData}, the obtained results are described in Section~\ref{Res}. A discussion is made in Section~\ref{Disc}, with a summary in Section~\ref{Summ}.


\section{Observational Data and Analysis Procedure}
\label{ObsData}

In this study of AR evolution, we used unprecedented, uninterrupted photospheric vector magnetic field observations obtained by the \textit{Helioseismic and Magnetic Imager} (HMI, \citealt{schou2012}) aboard Solar Dynamics Observatory. HMI observes the full solar disk in the Fe {\sc i} 6173\AA~spectral line with a spatial resolution of 0''.5/pixel. Filtergrams are obtained at six wavelength positions about the line center to compute Stokes parameters I, Q, U, and V. These are then reduced with HMI science data processing pipeline \citep{hoeksema2014} to retrieve the vector magnetic field using Very Fast Inversion of the Stokes Vector algorithm \citep{borrero2011} based on the Milne-Eddington atmospheric model. The inherent $180^\circ$ azimuthal ambiguity is resolved using the minimum energy method \citep{metcalf1994,leka2009}. Finally, the projection effects in the field components in native AR patch are corrected by transforming them to disk center using Cylindrical Equal Area (CEA) projection method \citep{calabretta2002}. These field components $(B_r, B_\theta, B_\phi)$ in heliocentric spherical coordinates are provided as \texttt{hmi.sharp.cea\_720s} data product, which are approximated to $(B_z, -B_y, B_x)$ in Heliographic Cartesian coordinates for ready use in various parameter studies \citep{gary1990}. Because the projection effects of the ARs situated beyond $\pm40^\circ$ longitude severely underestimate the field values and could not be recovered by disk transformation; we consider the AR's disk passage within this limit for their evolution study of net vertical current. 

Then, we derive the vertical component of electric current density and its decomposed components to twist/chiral and shear terms \citep{zhangh2001}
\begin{eqnarray}
J_z &=& \frac{1}{\mu _0}\left( \frac{\partial B_y}{\partial x}-\frac{\partial B_x}{\partial y} \right) \nonumber \\ 
&=&\frac{B}{\mu_0}\left( \frac{\partial b_y}{\partial x}-\frac{\partial b_x}{\partial y} \right)+\frac{1}{\mu _0}\left( {b_y}\frac{\partial B}{\partial x}-{b_x}\frac{\partial B}{\partial y} \right) \nonumber\\ 
&=&Jz_{ch}+Jz_{sh}
\label{Eq_curr}
\end{eqnarray}
where $b_x$, $b_y$, $b_z$ are components of the unit magnetic field vector $\mathbf{b}=\mathbf{B}/B$ with total field strength $B=|\mathbf{B}|$. The partial derivatives are approximated by finite difference scheme employing three-point Lagrangian interpolation procedure (See \citealt{vemareddy2017d} for more details). As the names suggest, the decomposition of the current density (${\mathbf J}$) is not for twisted and sheared magnetic field. ${J}_{ch}$ is connected to helicity term or force-free field but ${J}_{sh}$ is not because part of the current is perpendicular to the magnetic field (cross-field current). $Jz_{sh}$ actually, provides a quantitative description of the relationship between the magnetic shear and gradient of the magnetic field and can be written in a form $\frac{1}{{{\mu }_{0}}}\cos (\theta ){{\left( \nabla B \right)}_{\bot }}$ where $\theta$ is the shear angle of the photospheric transverse magnetic field, ${{\left( \nabla B \right)}_{\bot }}$ is horizontal gradient of magnetic field \citep{zhangh2001}. This term connects with the heterogeneity and orientation of magnetic field. Here field strength gradients play main role because cosine of $\theta$ ranges between -1 and 1.      

The net vertical flux and current are then deduced by summing the pixels over a particular polarity as given by $\!\!\Phi\!\!=\sum\limits_{N\,pix}{B_z}$ and  $I=\sum\limits_{N\,pix}{J_z}$ respectively. The net current in a given polarity flux ($I_S$ in south ($B_z<0$), $I_N$ in north ($B_z>0$), similarly the decomposed currents $I_{ch}$, $I_{sh}$) has contribution from both positive and negative currents. 

In order to have insight on the contribution of each signed current in the neutralization of the current in each polarity, we further separate the net current in each polarity into positive $(+)$ and negative $(-)$ components. The individual net currents in each polarity always have a systematic evolution and thus provides insights on their plausible generation mechanisms. We then evaluate the absolute ratio of these individual currents, as dominant and non-dominant net current in each polarity for the evaluation of the absolute ratio of direct-current (DC) and return-current (RC) \citep{torok2014}. Here, the dominant or non-dominant with respect to the net current value of positive (or negative) current over its counter signed net value in a given magnetic polarity. The sign of the dominant current in a given magnetic polarity gives the chirality of the AR. With this ratio, it is easy to monitor their relative time evolution in concurrent with the flux evolution. The ratio will always have a value greater than one in magnitude and a value of one indicates neutralized currents in a polarity. Deviation from unity level signifies non-neutralization and is a special case of non-potentiality. 

\begin{figure*}
	\centering
	\includegraphics[width=.99\textwidth,clip=]{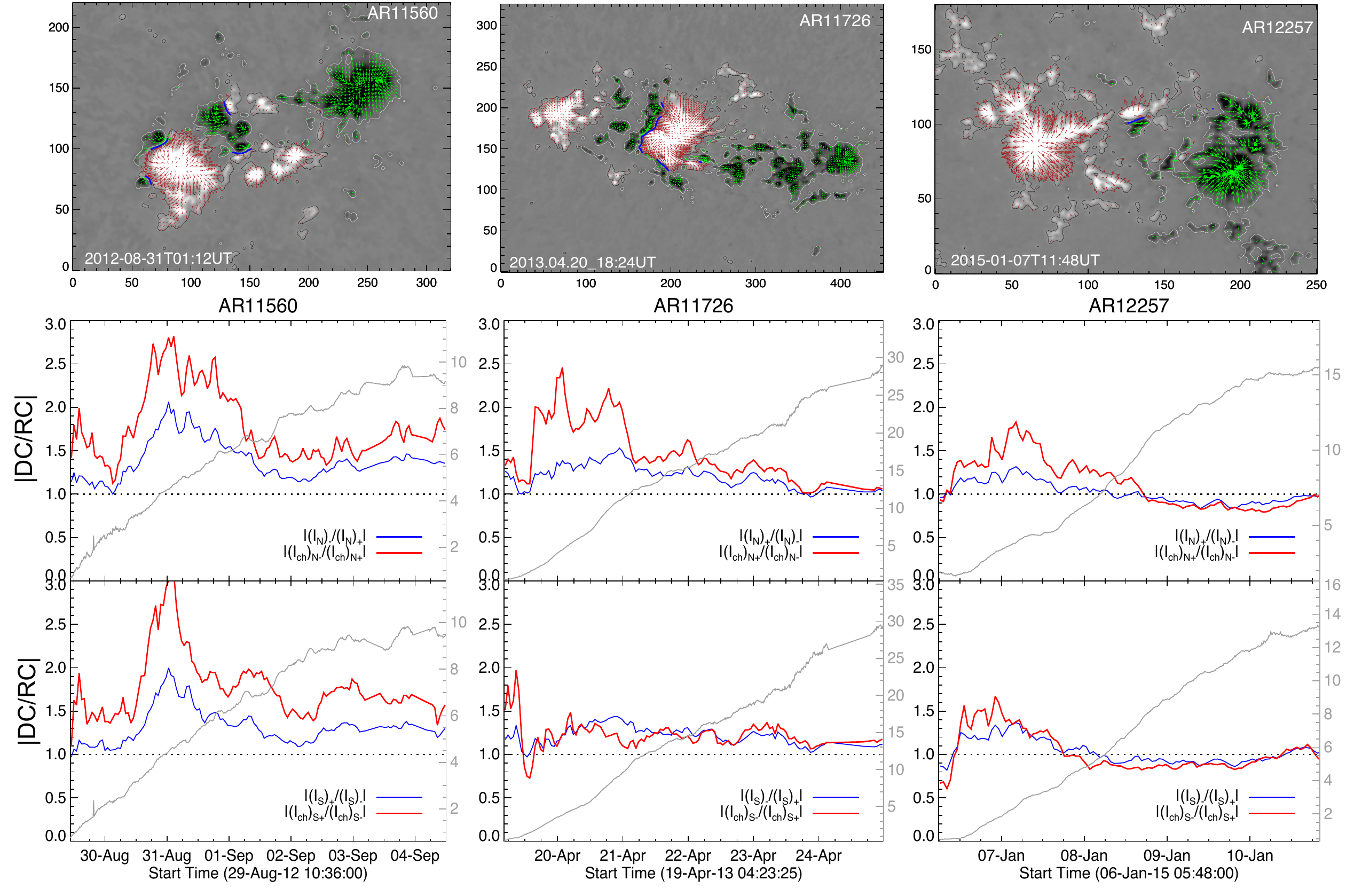}
	\caption{Evolution of $|DC/RC|$ in north ($B_z>0$, top panel) and south ($B_z<0$, bottom panel) in emerging ARs. The ratio is also plotted for $Jz_{ch}$ (red curve) in respective panels. Net magnetic flux (in units of $10^{21}$ Maxwell) is plotted in grey with y-axis scale on right of each panel. Horizontal dotted line marks current neutralization level of unity. Note that $|DC/RC|$ for  $Jz_{ch}$ (volume currents) is dominant than for $Jz$ indicating the additional contribution of $Jz_{sh}$ in the NCN. A sample vector magnetogram is displayed in the corresponding first row panels with traced SPILs (blue curves). Transverse field vectors are shown with red (green) arrows in north (south) polarity regions. Axis units are in pixels of 0''.5 size.   }
	\label{Fig2}
\end{figure*}

From Equation~\ref{Eq_curr} of vertical current decomposition, field strength gradients are known to contribute opposite signed currents to that contributed by field direction in each polarity regions. The combined result is more net current neutralization \citep{vemareddy2017d}. To know the extent of this contribution, we estimate $|DC/RC|$ for $I_{ch}$, in a similar way, by removing current contributed from field strength gradients ($I_{sh}$). 

For observationally reliable values, we consider pixels having threshold values $|B_{tot}|>200$\,G in all above computations. These strong fields also exists with uncertainties due to the field measurements and inversion procedure. The inversion products are given with estimated errors in the vertical component varying by about 40 G and in horizontal components by 80 G. As a test case, we propagate these uncertainties through Equation~\ref{Eq_curr} \citep[see details in][]{vemareddy2015c} by the equation 
\begin{equation}
\left(\delta J_z\right)_{i,j} = \frac{1}{\mu_0}\left(\frac{ \sqrt{\left(\delta B_x\right)_{j-1}^2+ \left(\delta B_x\right)_{j+1}^2 + \left(\delta B_y\right)_{i-1}^2
		+\left(\delta B_y\right)_{i+1}^2}}{2\Delta x}\right)
\end{equation}
where i, j refer to pixel indices in x and y direction, respectively. An equal spacing grid size in both x and y-directions is assumed in arriving at the above expression. Taking an average error  $\delta B_x=\delta B_y=40$G in a typical distribution of $n=10^4$ pixels, we found that the range of uncertainty ($\sqrt{n}\times \delta J_z dx dx$) of a signed net current in a given polarity can never be larger than $0.1\times10^{12}$A. Here $dx$ is HMI pixel size of $0''.5$ . Therefore, our estimation of $|DC/RC|$ can have a maximum error limit of 0.14. As we will see in the later section, this uncertainty is very small compared to the range of $|DC/RC|$ evolution in CME and flare producing ARs. 

We also traced the sheared PILs in every one-hour snapshot by an automated procedure similar to the one applied for tracing PILs of strong vertical field gradients \citep{mason2010} and applied in a statistical study \citep{Vasantharaju2018}. The sheared PIL is a segment with shear angle (difference between directions of observed transverse fields $\mathbf{B}_o$ and potential transverse fields $\mathbf{B}_p$ and is given by,  $\theta_{sh}= cos^{-1}(\mathbf{B}_o \cdot \mathbf{B}_p / |\mathbf{B}_o|| \mathbf{B}_p|)$)  greater than $45^\circ$.  In this procedure, we smooth the vertical magnetic field ($B_z$) image to a smoothing factor of 8 pixels (4 arcsec) and identified the zero Gauss contour. Then the potential magnetic field is calculated from the strongly smoothed magnetogram and the shear map is generated from the computed potential and the observed transverse field. On applying the thresholds of the strong observed transverse field ($>300$ G) and strong shear angle ($>45^\circ$) to contour segments, the strong field and strong shear PILs are identified. The summation of these PIL segment lengths gives the total length of SPIL present in the AR.

\begin{table*}
	\caption{Details of the Active Regions in this study}
	\begin{threeparttable}
	\centering	
	\begin{tabular}{lllllll}
		\hline\hline
		s. no & NOAA AR &  Latitude \tnote{a}  & Duration \tnote{b} & Chirality \tnote{c}& activity \tnote{d}  &   $|DC/RC|$ \tnote{e} \\
		& & & & & &   (min-max) \\
		\hline\hline 
		1 &11560 & N3    & 29 Aug--4 Sep, 2012   &  negative   & M-C flares \& CMEs    & 1.0--2.0   \\
		2 &11726 &  N13 & 19--25, April, 2013      &  positive     &  M-C flares \& CMEs  & 1.0--1.5 \\
		3&12089 & N18  &  11--15 Jun, 2014        & positive       &  M-C flares \& CMEs   & 1.0--1.9  \\
		4 & 11460 & N16& 17--22 Apr, 2012	 & 	positive	 & M-C flares \& CMEs  &  1.0--1.3 \\
		5 &12257 & N05  &   6--10 Jan, 2015         &  positive		& M-C flares \& CME  & 1.0--1.3  \\
		\hline 
		6&11429 &	N17			& 5--10 Mar, 2012    & negative		  &  X-C flares \& CMEs	 & 1.6--2.0 \\
		7 &11967 & S12	 & 31 Jan--6 Feb, 2014	&	negative  &	 M-C flares \& CMEs    	&  1.0--1.7  \\ 
		8& 12158 & S21  & 7--11 Sep, 2014     &  negative         & X-C flares \& CMEs     &	1.3--1.6 \\
		9&12371 &	N12	& 18--25 Jun, 2015	& negative		  &  M-C flares \& CMEs		  &  1.1--1.4  \\			
		10 & 12403 & S13 & 21--26 Aug, 2015	& 	positive	& 	M-C flares \& CMEs   &		1.0--1.2		\\
		\hline 
		11 &11166 &N11	&  7--11 Mar,2011   &  positive		 &    X-C flares	&	    1.0--1.1   \\
		12 & 11190 & N12 & 11--16- Apr, 2011	 & 	negative		& 	M-C flares		&  	1.0--1.1		\\
		13 &11928 & S15  & 16--21 Dec, 2013 &  positive     	  &   C-flares   	&  	1.0--1.1 \\   
		14 &12192 & S14	 &  21--26 Oct, 2014 	&  negative		   &  X-C flares  & 	1.0--1.1	 				   \\
		15 & 12422 & S20 & 23--29 Sep, 2015	& 	negative	& 	M-C flares			&	1.0--1.2				\\  
		\hline
		16 &11072 & S16  & 20--27, April 2010  &  negative      & quiet    &   		1.0--1.2    \\
		17 &11987 &  N07   & 25--26 Feb, 2014 & positive &      quite       	&  1.0--1.05  \\
		18 & 12082 & N14   &    6--10 Jun, 2014    &  negative  &   quiet    	&  1.0--1.1\\
		19& 12414 & S10		& 10--12 Sep, 2015	 & negative		& quiet			&  1.0--1.1 \\
		20 & 12600 &  N13   &10--14 Oct, 2016   & positive  &    quiet  	&  1.0--1.1\\  
		\hline\hline		
	\end{tabular}
\begin{tablenotes}
	\item[a] Latitudinal position of the AR; north (N)  or south (S)
	\item[b] period of the AR in this study during its disk transit within $\pm$40$^\circ$ longitude
	\item[c] twisted nature or handedness (left or right) of the dominant AR flux system   
	\item[d] observed GOES X-ray flares ($>$C1), LASCO-C2 CMEs
	\item[e] maximum uncertainty is 0.14 as deduced in Section 2 
\end{tablenotes}
	\label{tab:Tab1}
\end{threeparttable}
\end{table*}

\section{Results}
\label{Res}
The distribution of vertical current and its components are examined in all the studied ARs. In all the cases, we notice that the $Jz$ and $Jz_{ch}$ have similar distribution in terms of sign and morphology. However, $Jz_{ch}$ and $Jz_{sh}$ vary both in sign and magnitude. As exemplary cases, these distributions are shown in three ARs in Figure~\ref{Fig1}. Signed maps in the last column panels show distribution of currents that are having opposite (same) sign of $Jz_{ch}$ and $Jz_{sh}$ and are indicated by -1 (1)  value. These pixels of opposite sign current distribution are varying between 60-65\% in the studied AR cases. This kind of distribution indicates that the gradients of field strength contributes to opposite signed current to that contributed by magnetic field direction in the vertical component of current, which was first found by \citet{vemareddy2017d}. Further, the $Jz_{sh}$ seem to show much-higher spatial-frequencies than $Jz_{ch}$. This may be considered to be consistent with small-scale structures present in sunspot penumbrae. The investigations by \citet{zhangh2010} also support our observations. They found that the individual fibrils are dominated by the current density component caused by magnetic inhomogeneity, while the large-scale magnetic region is generally dominated by the twist component of the electric current density. \citet{vemareddy2017d} examined this concept of spacial frequencies by smoothing the magnetograms with different spacial windows. Because of large-scale structure, the $Jz_{ch}$ retains the spatial patterns even after smoothing while $Jz_{sh}$ smears the patterns due to higher spatial frequencies related to small-scale structure. 

Then, we observed the time evolution of net chiral (or twist) current ($I_{ch}$) in a polarity follows the trend of net current $(I)$ at a higher magnitude level. Depending on the sign of $I_{sh}$, the DC or RC of $Jz$ will be differently deviated from the DC or RC of $Jz_{ch}$. We evidenced the sign of $I_{sh}$ is mostly (2/3) negative to that of $I_{ch}$ in a given polarity, consequently decreasing the DC value (than that of RC) of $Jz$ resulting in higher degree of net current neutralization (NCN) than the net twist current neutralization. In order to make this clear, we show this difference of neutralization for both $I_{ch}$ and $I$ in each polarity of an AR. 

In the following, we study the evolution of $|DC/RC|$ in 20 different ARs of different coronal activity. Most of these regions are well studied previously in different contexts. For clarity, we analyzed all of them, and then divided into groups of their evolution phase and the nature of activity. The emerging ARs consists of emerging bipolar regions with sub-photospheric twist and complex interaction whereas in the post emerged cases, the AR emergence phase is missed and the evolution consists of flux fragmentation, diffusion, shear motions and/or flux emergence in the pre-existing flux. As we already know that the ARs in decaying phase are sometimes very active (e.g., AR12371) and we search such cases to include here. On the other hand, there are ARs producing flaring activity associated with/without CMEs, the so called confined/eruptive ARs. This classification has implications to the space-weather exploring the conditions of an AR to be CME productive.  In principle, the sample can be more than 20, but restricted to put up results more clearly. Flares are classified by disk integrated GOES X-ray flux\footnote{provided at \url{https://www.solarmonitor.org}}. Coronal mass ejection (CME) is a large scale expulsion of magnetic field and plasma from the AR disturbing the outer corona at least to the LASCO-C2 field-of-view extent. To associate the CMEs with the studied source ARs, simultaneous coronal observations are scrutinized in association with the LASCO-C2 images\footnote{obtained from \url{http://cdaw.gsfc.nasa.gov/CME\_list/UNIVERSAL/} } \citep{yashiro2004}. From this qualitative information, we noted the type of coronal activity in relevance to the net current evolution in the AR. A summary of the AR cases in this study is given in Table~\ref{tab:Tab1}. As providing the details of the individual events in all 20 ARs is spatially prohibited, besides referring to the earlier studies, we noted the timings of prominent CME/flare events in the text.

\begin{figure*}
	\centering
	\includegraphics[width=.99\textwidth,clip=]{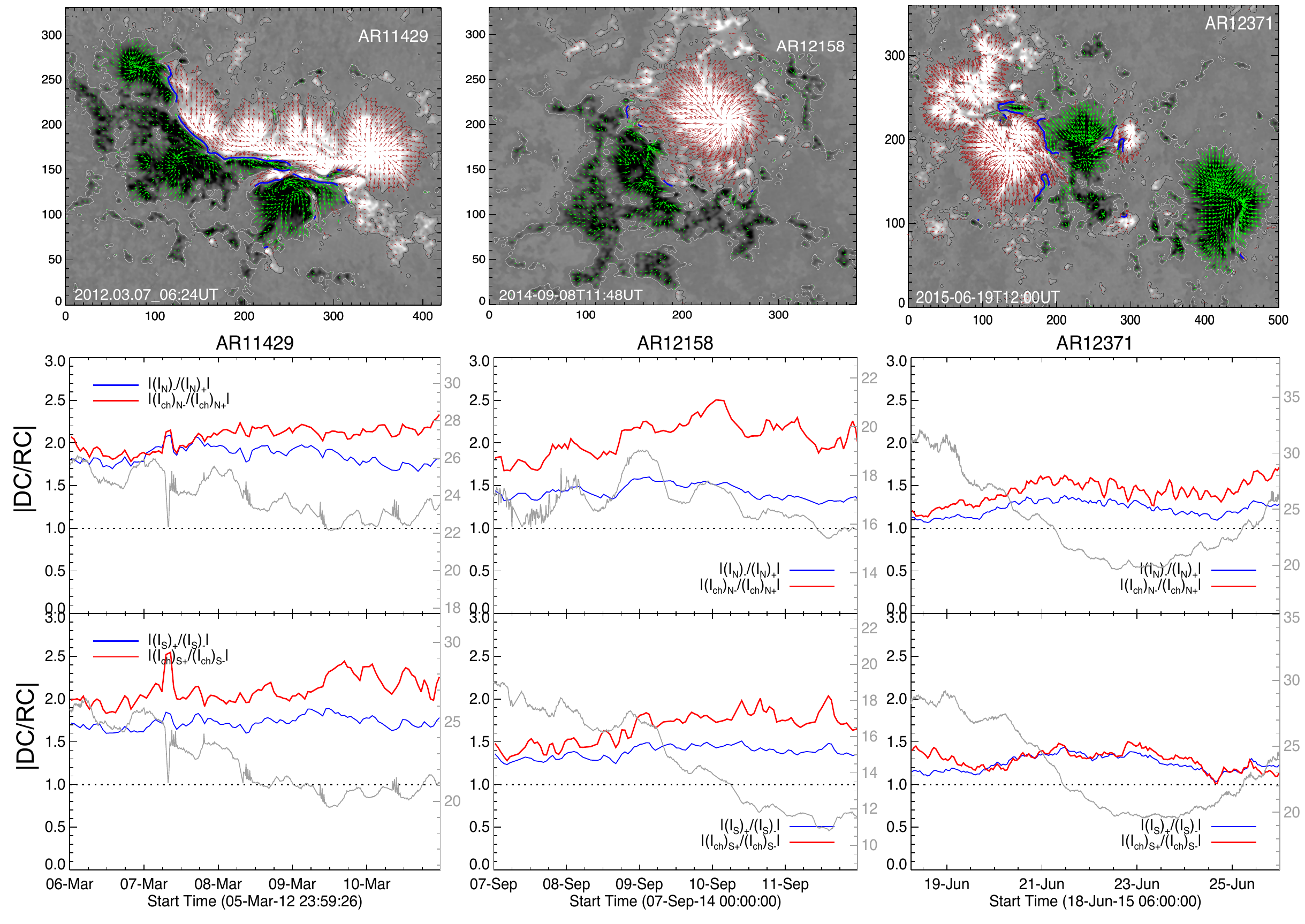}
	\caption{Same as Figure~\ref{Fig2}~but for post emerged ARs producing CMEs and associated flares.}
	\label{Fig3}
\end{figure*} 

\subsection{Emerging ARs}
We choose set of five emerging ARs with CME and flaring activity and studied the NCN with time. In Figure~\ref{Fig2}, time evolution of NCN is plotted only for three AR cases. The $|DC/RC|$ is shown both for positive and negative magnetic polarity in separate panels. Horizontal dotted line indicates the NCN of unity referring to equal values of DC and RC in a magnetic polarity. The net flux of each polarity is also plotted in the same panels with y-axis scale (in units of $10^{21}$ Maxwell) on the right side. A sample vector magnetogram is depicted in the corresponding first row panels with traced SPILs (blue curve). In all five AR cases, the $|DC/RC|$ starts near a value of one and increases rapidly with the fast flux emergence during about first 20 hours. We noticed that this fast emergence is given by appearance of compact bipolar regions having SPILs at the interface of opposite polarities. This phase is then followed by separating motion of bipolar regions becoming isolated bipolar regions. As a result, ARs become less compact without SPILs as the interface of opposite polarities diminishes. We see that the $|DC/RC|$ decreases from the peak value (1.5-2) of highest compact phase to the unity of well separated polarity regions. As we can notice that all the ARs have higher value of $|DC/RC|$ for $Jz_{ch}$ than for $Jz$.

All five AR cases (serial no 1-5 in Table~\ref{tab:Tab1}) follow the profile of the NCN exhibited by simulated emerging bipolar AR reported by \citet{torok2014}. The simulation results (their Figure 4) showed the development of strong deviation from NCN simultaneously with the onset of significant flux emergence into the corona, accompanied by the development of substantial magnetic shear along the AR PIL. After the region has formed and flux emergence has ceased, the strong magnetic fields in the regions center are connected solely by DCs and the total DC is several times larges than the total RC.   

\begin{figure*}[!htp]
	\centering
	\includegraphics[width=.99\textwidth,clip=]{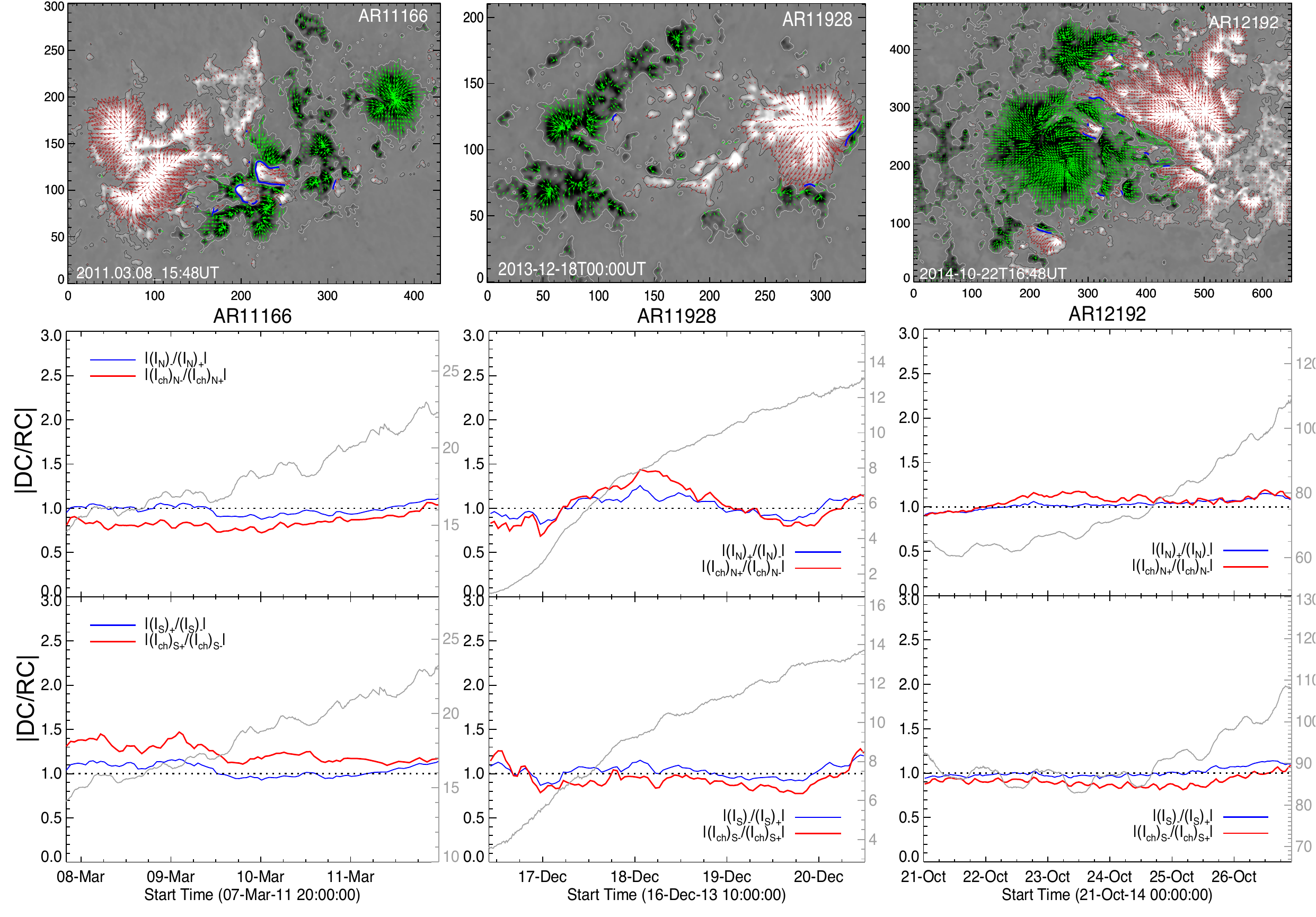}
	\caption{Same as Figure~\ref{Fig2} but for flaring ARs.}
	\label{Fig4}
\end{figure*}

The AR 11560, 11726 are source regions of C-M flares and CMEs during the present observation window. A prominent CME occurred in AR 11560 at 1:00UT on September 2, 2012, which is after the peak phase of $|DC/RC|$ value. In the AR 11726, the coronal eruptions are seen at 20/07:40UT, 22/06:00UT. Owing to the opposite chirality, the AR 11560 have opposite signed DC value than that in AR 11726. These two cases are studied in the context of helicity and energy flux injection in \citet{vemareddy2015a}. The breakdown of NCN in these two emerging ARs appears to have correspondence with the CME occurrence, in agreement with the proposed CME initiation mechanisms based on flux emergence \citep{leka1996}.  A similar $|DC/RC|$ evolution follows in ARs 11460, 12089, 12257 however, the activity is limited to C-M flares without prominent CMEs.          

\subsection{Post emerged ARs}
In Figure~\ref{Fig3}, the time evolution of $|DC/RC|$ in post emerged ARs is plotted. As pointed already, the ARs whose emergence phase is not observed are referred as post emerged ARs. In the post emergence phase, the bipolar regions are well separated however in some ARs of multi-polar regions, the separation leads to interaction among other polarities causing the development of interfering regions with PILs. In the case of interaction is accompanied by converging and shearing motion between opposite polarities, the interface develops to have shared PIL as $\delta$-configuration. The AR 11429 has a larger interface of SPIL ($\approx130$Mm) as seen in the vector magnetogram panel and maintains the $|DC/RC|$ value above 1.5 throughout the evolution period. While maintaining high degree of non-neutralization, the AR 11429 produced 7 M, 3 X-class flares, some of which are accompanied by CMEs \citep{LiuLijuan2016}. To be noted, the CMEs on 05/1:55UT, 07/00:05UT, 09/03:40UT, 10/17:30UT are very severe events.  This AR is also studied in \citet{sunx2015} in the activity context and \citet{YangLiu2017} in the NCN context. 

The AR 12158 has large rotating south polarity sunspot surrounded by plage type north polarity distribution comprising the SPIL. The NCN is near to 1.5 in the observed period of evolution and have one M-class and one X-class flare associated with CMEs \citep{vemareddy2016b}. At the time of the observed two major CMEs (08/23:00UT, 10/17:15UT), the degree of non-neutrality is obove 1.5. The AR 12371 has a following bipolar region with main SPIL interface and an isolated leading south polarity. In our calculation of NCN, the leading south polarity is included which may lower the estimated $|DC/RC|$. The $|DC/RC|$ is well above unity to a maximum of 1.3 and the observed activity is CME associated M-flares (18/16:25UT, 21/02:00UT, 22/17:39UT,24/15:12UT, 25/08:02UT) as studied in \citet{vemareddy2017b}.  Note that the net magnetic flux in these cases is decreasing over time to indicate the decay phase of the AR. Similarly, the ARs 11967, 12403 evolve with strong deviation of NCN and generates C-M flares some of them are accompanied by CMEs.

\begin{figure*}
	\centering
	\includegraphics[width=.99\textwidth,clip=]{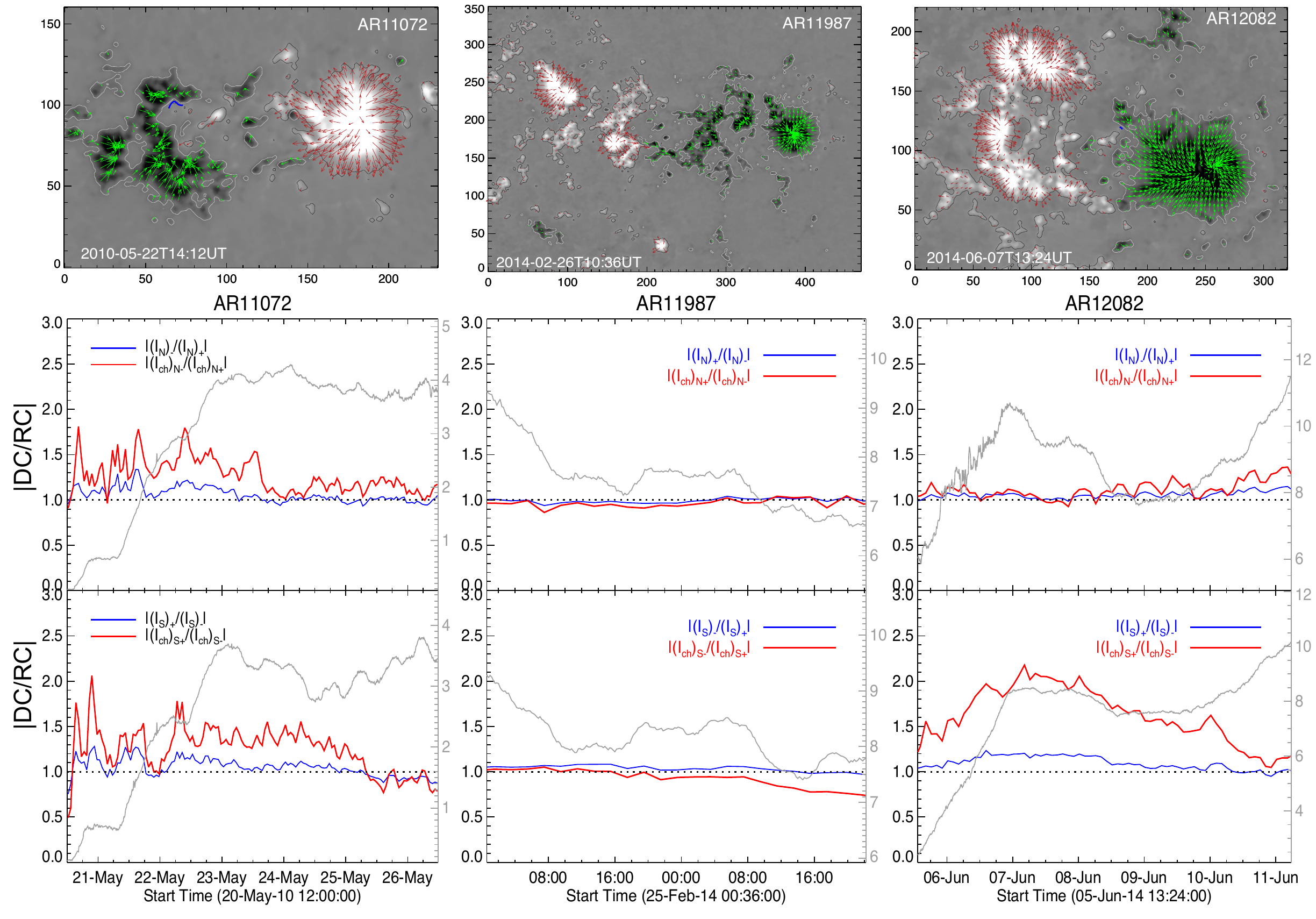}
	\caption{Same as Figure~\ref{Fig2} but for quiet ARs.}
	\label{Fig5}
\end{figure*}

\subsection{Flaring ARs}

Differently, we choose another set of ARs having only flaring activity. In Figure~\ref{Fig4}, the time evolution of $|DC/RC|$ is plotted for three of these ARs. The current neutralization oscillates near unity in AR 11166 despite flux emergence in the post phase of the AR birth. The observed activity is related to flares including X1.5 at 23.13 UT on March 9, 2011 \citep{vemareddy2014a}. The SPILs are in the form of segments due to embedded north polarity in the south polarity region. The AR 11928 is an emerging one but $|DC/RC|$ undulates about unity level unlike the cases in the group of emerging ARs. No prominent SPIL segments are noticed in the AR polarity distribution. As studied in \citet{vemareddy2017a}, the chirality changes from positive to negative over the observed period of evolution generating C-flaring activity. The evolution in AR 12192 is very near to unity satisfying the NCN throughout the evolution. Despite being super AR, no long SPIL is present except small segments as noticed in the vector magnetogram panel. Probably of these reasons, the AR is flare productive including major X-flares on 22/14:02UT,24/21:07UT, 25/16:55UT, 26/10:04UT,27/14:12UT none of them are associated with CMEs \citep{LijuanLiu2016,Amari2018}. Our $|DC/RC|$ profile is very much agreement with \citet{YangLiu2017}, although derived without applying any masking procedure. Similarly, the derived NCN profiles for AR 11190, 12422 imply near unity evolution related to flaring activity without CMEs.

\subsection{Quiet ARs}
ARs without flaring activity greater than C-flares and CMEs are taken for the study of NCN evolution. Three of such ARs are shown in Figure~\ref{Fig5}. In AR 11072, we see fluctuations of $|DC/RC|$ deviating from neutrality during the emergence phase till mid of May 23, which diminishes to neutrality level as the bipolar regions separates without SPILs. Different from emerging ARs discussed earlier, the quiet activity nature could be likely due to lack of twist in the emerging field. The evolution of NCN in AR 11987 follows almost unity. In AR 12082, the evolution of NCN follows unity in north polarity but deviates significantly from unity in south polarity. A small difference in degree of NCN between both polarities may arise (also in \citealt{YangLiu2017}) due to various errors in horizontal field measurements (related to observational sensitivity) and thereby the numerical differentiation. Importantly, the difference would be magnified by division operation in the ratio of DC and RC. We can have flux imbalance that is related to plage regions, or individual polarities because the non-neutralized currents comes only from SPIL vicinity. Since we show the NCN in each polarity, the imbalance is not a concern at least to the present scope of the study. In those cases, it may require to choose exact region of SPIL although considering entire AR yields expected evolution trend in most of the ARs of our sample.

\section{Discussion}
\label{Disc} 
Non-neutralized currents arise in the vicinity of SPILs amid opposite polarity regions in close proximity \citep{georgoulis2012}. The numerical study of \citet{torok2003} first showed that as soon as shearing motions are generated in the PIL area, non-neutralized currents inevitably build up in the corona. In their study, shear motions are generated by vortex flows and the total non-neutralized current is proportional to distance between the opposite flux concentrations. These studies are extended further to acknowledge the fact that the breakdown of electric current neutralization occurs exclusively along SPILs \citep{torok2014,YangLiu2017}. 

\subsection{Different views of direct and return currents}
In all the AR cases, we see that the $|DC/RC|$ is more for $Jz_{ch}$ than for $Jz$. It proves that the current due to field strength gradients $Jz_{sh}$ have opposite sign distribution dominantly (60-65\%) to that of field direction $Jz_{ch}$, substantiating the result first found in AR 11158 by \citet{vemareddy2017d}. This \citet{parker1996} flux tube view of current neutralization is explained differently by \citet{dalmasse2015} with MHD simulations. In their view, instead at the interface of flux tube and the field free plasma, the return current is also a volume current belonging to flux bundle forming a more or less thick shell between the direct current at the core of the flux tube and the surrounding potential field (or field-free plasma if the flux bundle is embedded in a field-free plasma). From this setup, they inferred that the existence of the return current is purely a geometric effect and is due to the presence of differential twist (variation of twist as a function of the distance to the flux bundle axis) in the flux bundle which is necessary to keep the flux bundle confined. 

Moreover, in the zero-$\beta$ MHD simulations of \citet{dalmasse2015}, the imposed motions along the PIL generate force-free net current. However, this is not the scenario on the photosphere, where the Lorentz-force driven shear generates cross-field currents too \citep{georgoulis2012} as the observations implied here. In the photosphere, magnetic pressure tends to expand the flux tube against the twist that keeps the flux tube confined. 

Our analysis assumes that return/sheath current is a surface current related to cross-field current and coming partially from ${\mathbf{J}_{sh}}$ \citep{georgoulis2004b, georgoulis2012}. Strictly, this does not prove that all return currents are pressure-driven cross-field sheath-like shells. This only shows that the force-free-like currents as considered in low/zero-beta coronal MHD-models only reveal parts of the currents that are present in the photosphere. Indeed Figure~\ref{Fig1}(left column), does show the existence of both DC and RC in $Jz_{ch}$, as seen in low/zero-beta models. So the result is that $I_{sh}$ contributes to the total return currents.  With this only similarity, the models are said to be incomplete for not having the real photospheric scenario of the currents.

The decomposition of the net current implies the importance of $I_{ch}$ in the coronal processes like flares/CMEs, as supported by low/zero-beta MHD models. $I_{sh}$ comes from in-homogeneous pattern of field-strength and therefore its evolution bears no trend. Higher in the corona, the field distribution becomes smooth, and $I_{sh}$ is negligible leaving only force-free current $I_{ch}$ \citep{vemareddy2017d}. 

While Parker's neutralized currents require in-situ storage and release of energy, the electric circuit models \citep{melrose1991,melrose1995}, on the other hand believe that the currents have to close at the base of the convection zone where magnetic fields are generated. In this scenario, the currents have to emerge from the solar interior with the emergence of the magnetic flux. Then the magnetic loops are electric circuits which have a strong inductive coupling between coronal flaring volume and the convection zone. In both the models, we anticipate the importance of $I_{sh}$ in generating solar activity in the ARs. In case $I_{sh}$ is strong, the Lorentz-force may generate shear motions in the vicinity of the PIL. The shear generating Lorentz-force further has physical connection to the non-neutralized currents (See \citealt{georgoulis2012}). Although this exercise is worth to study, we believe that the Lorentz-force related to $I_{sh}$ would be small. It is clear from Figure~\ref{Fig1} that the $I_{sh}$ comes from almost everywhere in the AR, so it is unlikely that the $I_{sh}$ play significant role in the active events like flares/CMEs. However, an organized distribution of $J_{sh}$ may have strong Lorentz-force. Based on SOT/HINODE vector magnetic field observations, \citet{georgoulis2012} had already showed that the shear generating Lorentz-force is significant in the ARs. Given the contribution of $I_{sh}$ in the net-current neutralization, our results favour the concept of shear-generating Lorentz force introduced by \citet{georgoulis2012}.

\subsection{SPIL length and non-neutralized current}
Following the connection between the magnetic shear and the non-neutralized currents, the length of SPIL is expected to follow the non-neutralized current evolution in the AR. This property is observationally verified in AR 11158 \citep{vemareddy2015c}. The AR 11158 has a systematic evolution of the net current in the two sub-regions where the SPIL is due to predominant rotational and shear motions of opposite flux concentrations. It was seen that the net current evolution follows the time evolution of total length of SPIL in both of the sub-regions. 

\begin{figure*}
	\centering
	\includegraphics[width=.7\textwidth,clip=]{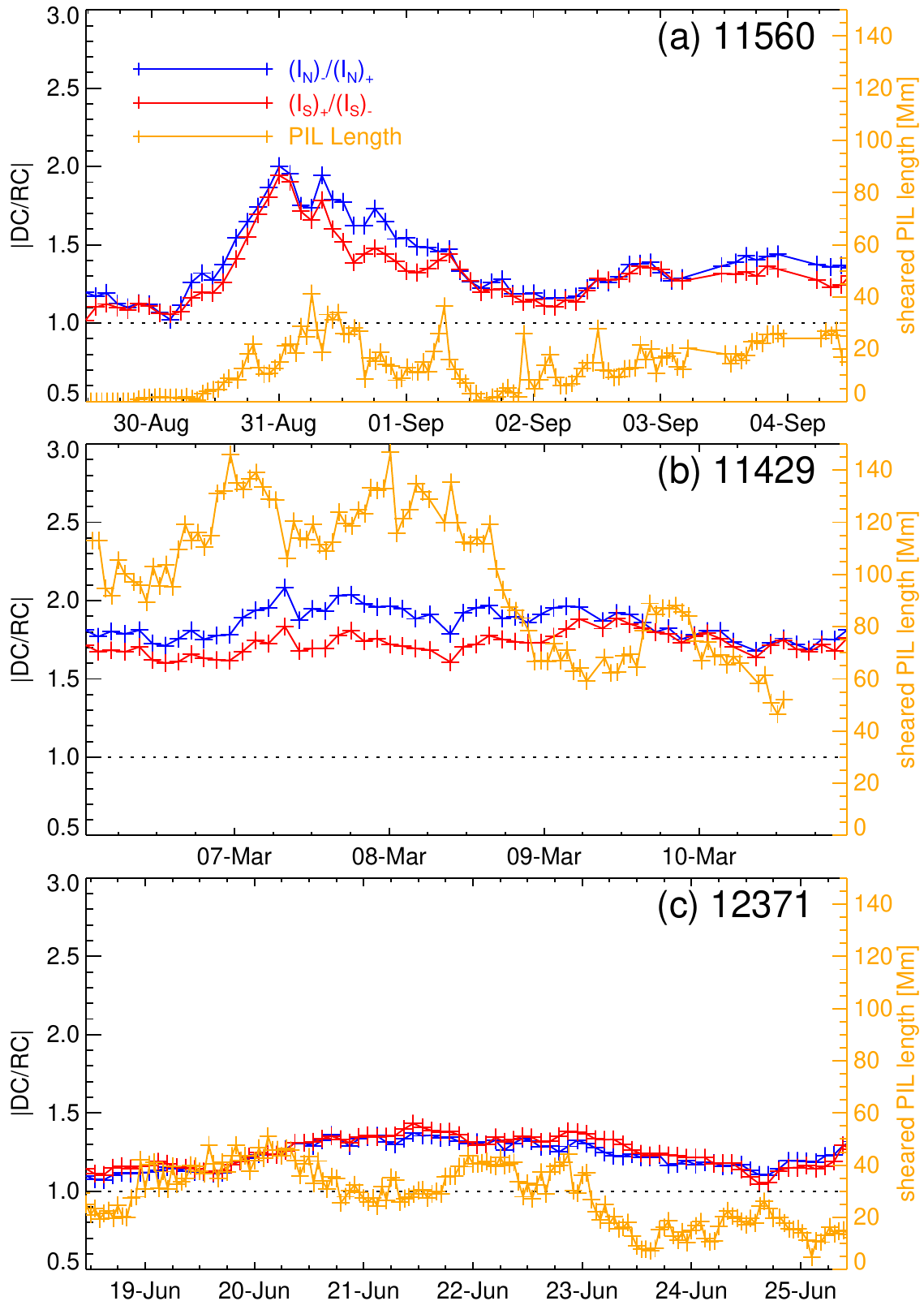}
	\caption{Relationship between SPIL length and $|DC/RC|$ in AR 11560, 11429 and 12371. Total length of all SPIL segments is plotted with time in the respective panels with y-axis scale on right side. Note that SPIL length correlates well (CC=0.69) with the non-neutralization in the AR 11560.   }
	\label{Fig6}
\end{figure*}

In Figure~\ref{Fig6}, we compare the profile of $|DC/RC|$ with the total length of SPIL segments in three ARs. The SPIL segments are traced by a systematic procedure as explained in section~\ref{ObsData}. Throughout the evolution right from the emergence, the total length of all SPIL segments correlates well (at a correlation coefficient of 0.69) with the non-neutralization profile in the AR 11560. A similar correlation is found by \citet{Vemareddy2019} between the SPIL length and the $|DC/RC|$ in the eruptive AR 12673.  The AR 11429, 12371 have long SPIL interface which have weakly correlated, at a correlation coefficient of 0.45, 0.29 respectively, with the $|DC/RC|$ profile. It means that the non-neutralized currents spread in a broader region about PIL. This could likely be the case in ARs containing large interface of opposite polarities where the shearing regions extend into the opposite polarities. Then the spread of sheared region along PIL may not represent the length of SPIL interface. This aspect of SPIL is worth of further detailed investigation in few more AR cases.

Interestingly, we find that the total SPIL length in AR 12192, 11166 is significantly large (60-150 Mm, 5-115Mm) during entire evolution period. It is in the form of small segments located all over the AR as shown in vector magnetograms in Figure~\ref{Fig4}. Obviously, the net-current in these ARs is large (orders of $10^{12} A$) and probably caused large flares but none of them associated with CMEs. However, the $|DC/RC|$ curve remains neutralized. We should point that the continuity of the SPIL is dependent on the threshold applied to trace the PIL. Given a lesser threshold ($<$300 G), the PIL can be continuous or will have larger segments through weak-field regions. However what matters is the magnetic shear information in the horizontal field components that in turn connects to the strong twisted structure along the PIL. Due to weaker horizontal field in regions below 300 G, the magnetic shear could be weak. To our knowledge, there exists no reports of flux rope structure above along the PIL in weak field regions less than 300 G. Therefore, our distinction of continuous nature of SPIL is linked to magnetic shear and twisted flux structure which is a key feature for the CME eruptions.

\subsection{Nature of magnetic shear along PIL}
We know that strongly twisted magnetic flux tubes and strongly sheared structures are candidates for explaining the intense flares and the associated flux rope eruption from AR. The segmented or localized distribution of magnetic shear indicates that the flares are relatively local phenomena, and the released energy is mostly converted into radiation and energetic particles. On the other hand, magnetic shear along a long continuous PIL interface may refer to coherently twisted magnetic structure, i.e., a flux rope, as require by many CME models \citep{amari2003a,torok2005}. Thus the CMEs are more global phenomena, and the energy mostly goes into mechanical energies through the ejection of magnetized plasma structures. Therefore, as opposed to general view, an intense flare may not necessarily be accompanied by a CME (e.g., \citealt{Feynman1994,Green2002,yashiro2005,WangYumin2007}) because the occurrence of a CME is substantially determined by the driving force of the inner core magnetic field and the confining force of the external overlying field. From this comparison, the magnetic shear can be distributed along a continuous PIL referring as global shear in the AR scale or it can be present along few segmented PILs known as localized shear.  The global shear case is a manifestation of strong twisted flux rope that may erupt under certain instability conditions. The local shear case may not form a twisted flux rope or even exists, the most flux in the AR acts as a confining environment (magnetic cage) for an eruption to takes place (see also the study of \citealt{vemareddy2017b, Amari2018}). From this observational analysis of the magnetic shear and the nature of the activity, we suggest that the confined or eruptive character of an AR is related to the nature of magnetic shear along the PIL.

\section{Summary and Conclusion}
\label{Summ}
We studied the connection of non-neutralized currents and the observed activity in 20 solar ARs. Net current evolution in the AR is derived using every 12 minute HMI vector magnetograms. For a more general view of the results, we separate the ARs according to their evolution phase, and the nature of activity and studied the degree of NCN. 

According to the flux rope models of CMEs eruptions \citep{zakharov1986}, the hoop force of the CME flux rope is proportional to square of the net current in the flux rope channel, thus the breakdown of net current neutralization implies to a form of Lorentz force development leading to stability loss and the CME eruption. The net current in a given polarity is algebraic sum of DC and RC, in particular their distribution may be relevant for CME/flare activity \citep{forbestg2010}.  The ratio $|DC/RC|$ is a measure of degree of net current neutralization due to magnetic shear and twist.

In the emerging ARs, the non-neutrality of the currents builds with the onset of flux emergence, following the relaxation to neutrality during the separation motion of bipolar regions. Accordingly, those emerging ARs are source regions of CMEs occurred at the time of higher level non-neutrality. ARs in the post phase of their emergence can be CME productive provided they have interacting bipolar regions with converging and shearing motions. In these cases, the net current evolves with higher level ($>1.3$) of non-neutrality. The net current evolution in flaring and quiet ARs is different. The $|DC/RC|$ in these cases varies near unity indicating neutralized currents in the magnetic polarities. These observations demonstrate that the CME eruptions are due to non-neutralized currents in the ARs, as proposed by \citet{YangLiu2017}. 

With the decomposition of the electric current, in all the AR cases, we see that the $|DC/RC|$ is more for $Jz_{ch}$ than for $Jz$. It implies that the current due to field strength gradients $Jz_{sh}$ have opposite sign distribution dominantly (60-65\%) to that of field direction $Jz_{ch}$. This further substantiates the result first found in AR 11158 by \citet{vemareddy2017d}. Apart from its role in current-neutralization, $I_{sh}$ can be the origin of shear-generating Lorentz-force, which further connected to non-neutralized currents and the solar activity in the ARs \citep{georgoulis2012}.

Owing to the fact that the non-neutralized currents arise in the vicinity of SPILs, the total length of SPIL segments follows the degree of NCN with a strong correlation in AR 11560. This correlation is positive but weak in AR 11429, 12371 that could be related to spread of sheared region about PIL. These findings are in support of a recent observational study of the SPIL length and the $|DC/RC|$ in strong eruptive AR 12673 \citep{Vemareddy2019}. In flaring ARs, the NCN is closer to unity even with total length of SPIL segments is large. Moreover, we notice that the SPIL is continuous in CME producing ARs in contrast to multiple small segments in flaring ARs which makes a clear difference in the magnetic shear distribution in the AR scale in which we refer as global shear in the former cases and local shear in the later cases. Our study of net current neutrality demonstrates the dividing line between the CMEs and flares with the difference being in global or local nature of magnetic shear in the AR.  

\section*{Acknowledgements}
The data have been used here courtesy of NASA/SDO and HMI science team. We thank the HMI science team for the open data policy of processed vector magnetograms. I thank the referee for insightful comments and suggestions. P.V is supported by an INSPIRE grant under the AORC scheme of the Department of Science and Technology. 

\bibliographystyle{mnras.bst} 

%
%

\bsp	
\label{lastpage}
\end{document}